\documentclass[conference]{IEEEtran}

\usepackage{xcolor}
\usepackage{epsf}
\usepackage{times}
\usepackage{epsfig}
\usepackage{graphicx}
\usepackage{epstopdf}
\usepackage{amsmath}
\usepackage{amssymb}
\usepackage{amsxtra}
\usepackage{here}
\usepackage{rawfonts}
\usepackage{times}
\usepackage{url}
\usepackage{cite}
\newlength{\aligntop}
\newlength{\alignbot}
\makeatletter

\IEEEoverridecommandlockouts
\begin{document}
\title{\huge Towards a {Consumer}-Centric Grid: A Behavioral Perspective\vspace{-0.4cm}}

\author{\authorblockN{ Walid Saad$^1$, Arnold Glass$^2$, Narayan Mandayam$^3$, and H. Vincent Poor$^4$} \authorblockA{\small
$^1$ Wireless@VT, Bradley Deparmtent of Electrical and Computer Engineering, Virginia Tech, Blacksburg, VA, USA, Email: \url{walids@vt.edu}\\
$^2$ Department of Psychology, Rutgers University, New Brunswick, NJ, USA, Email: \url{aglass@rci.rutgers.edu}\\
$^3$ Electrical and Computer Engineering Department, Rutgers University, New Brunswick, NJ, USA, Email: \url{narayan@winlab.rutgers.edu}\\
$^4$ Electrical Engineering Department, Princeton University, Princeton, NJ, USA, E-mail: \url{poor@princeton.edu}\vspace{-0.5cm}
 }%
   \thanks{This research is supported by the U.S. National Science Foundation under Grants CNS-1446621, ECCS-1549894, ECCS-1549900, and ECCS-1549881. Dr. Saad was a corresponding author.}
 }
\date{}
\maketitle

\begin{abstract}
Active consumer participation is seen as an integral part of the emerging smart grid. {Examples include demand-side management programs, incorporation of consumer-owned energy storage or renewable energy units, and active energy trading}. However, despite the foreseen technological benefits of such consumer-centric grid features, to date, their widespread adoption in practice remains modest. To shed light on this challenge, this paper explores the potential of prospect theory, a Nobel-prize winning theory, as a decision-making framework that can help understand how risk and uncertainty can impact the decisions of smart grid consumers. After introducing the basic notions of prospect theory, several examples drawn from a number of smart grid applications are developed. These results show that a better understanding of the role of human decision-making within the smart grid is paramount for optimizing its operation and expediting the deployment of its various technologies.\vspace{-0.2cm}
\end{abstract}

\section{Introduction}
\begin{figure}[!t]
 \begin{center}
 \vspace{-0.2cm}
  \includegraphics[width=8cm]{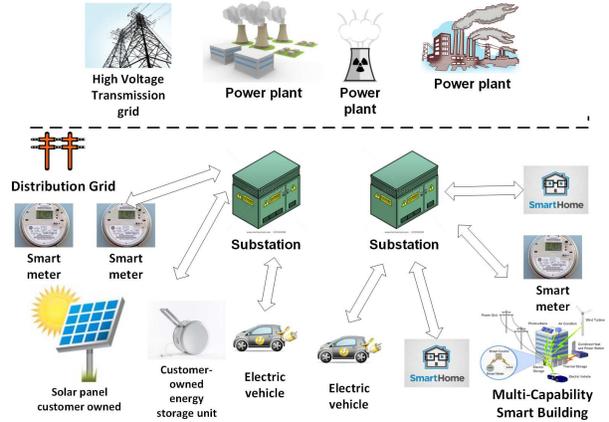}
 \vspace{-0.5cm}
   \caption{\label{fig:sm} A future smart grid with a heterogeneous mix of storage units, EVs, renewable sources, and other consumer-owned equipment.}
\end{center}\vspace{-0.9cm}
\end{figure}

\vspace{-0.1cm}
The electric power grid has undergone unprecedented changes over the past few years. The traditional, hierarchical and centralized electric grid has transformed into a large-scale, decentralized, and ``smart'' grid~\cite{SG00,SG01,SG02,FG00}. Such a smart grid is expected to encompass a mix of devices, as shown in Fig.~\ref{fig:sm}, that include distributed renewable energy sources, electric vehicles (EVs), and storage units that can be actively controlled and operated via a reliable, two-way communication infrastructure~\cite{SG00}. The effective operation of such a heterogeneous and decentralized system is expected to change the way in which energy is produced and delivered to consumers.

One key byproduct of the smart grid evolution is an ability to deliver innovative \emph{energy management} services to consumers~\cite{SG00,SG01,SG02,FG00}. Here, energy management refers to the processes using which energy is generated, managed, and delivered to consumers in the grid. For instance, demand-side management~(DSM) and demand response mechanisms will be an integral part of the smart grid. The primary goal of such programs is to dynamically shape and manage the supply and demand on the grid in order to maintain a desirable load over various timescales. Indeed, the design of optimized DSM and demand response protocols and associated pricing schemes has led to significant research in this area in recent years~\cite{HAM00,DSMDR00,DR02,MGA04,QZ00,DR03,DR04,DSMDR01,DSMDR02,FF05,GT04,FF03,FF04,CP04,CP00,CP01,CP02,CP08,DSMDR03,DSMDR04,DSMDR05,DSMDR06,DSMDR07,DSMDR08,DSMDR09,DSMDR10,DSMDR12,DSMDR13,DSMDR14,DSMDR15,DSMDR16,DSMDR17,DSMDR18}.

Moreover, in the smart grid, consumers will be able to individually own energy production units, such as solar panels, as well as storage devices in the form of EVs or small batteries. This can potentially transform every smart grid consumer into an independent energy production and storage source. Consequently, the possibility of \emph{energy trading} between such well-equipped consumers will undoubtedly become a reality in the next few years. Indeed, many recent works, such as in \cite{MGA,MGA04,GT06,HAM01,WI03,REN00,STOR00,STOR01,STOR02,STOR03,STOR04,STOR05,STOR06,STOR07,EW01,EW07,EW08,EW02,EW05,EW06,ET00,ET01,ET02,ET03,ET04,ET05,ET06,ET07,ET08,ET09}, have investigated the various challenges of such large-scale energy exchange, which include the development of optimized market mechanisms, the management of the grid operation, and the optimized exploitation of available consumer-owned storage and energy production units.

Realizing this vision of a distributed, sustainable, and consumer-centric smart grid will naturally face many challenges. On the one hand, although DSM programs (and related ideas) have been theoretically shown to yield important technological benefits to the grid, {their wide-spread deployment still remains insipid~\cite{FAHEY,FAHEY2,FAHEY3,FAHEY4,FAHEY6,FAHEY5,FAHEY7}}. On the other hand, the impact of energy trading on the smart grid operation and the realistic assumption that every consumer can become a producer of energy is still not well-understood. In addition, how to maximize the amount of energy that stems from renewable sources is yet another important challenge. Last but not least, the design of efficient dynamic pricing mechanisms that go hand-in-hand with DSM and energy trading schemes is seen as a critical enabler for most of the foreseen smart grid features.

{To widely deploy such grid features, one important challenge, among many others, is to properly incentivize consumers to actively participate in emerging grid features.} For example, without effective adoption of DSM schemes by consumers, power companies will not be able to reap the technological benefits of such load-shaping mechanisms. Similarly, the willingness of consumers to own and actively utilize EVs, storage units, or even renewable sources, is an essential milestone for the deployment of a truly sustainable, consumer-centric smart grid. {For example, the statistics in \cite{FAHEY6} show that installed solar generating capacity has increased from about 1000 MW in 2010 to more than 6000 MW in 2014. However, residential capacity has only increased from about 200 MW in 2010 to about 1000 MW in 2014. The reasons for this small growth are touched upon in \cite{FAHEY5} -- the upfront cost of \$21,000 - \$25,000 is more than most homeowners have to risk on what is still an uncertain venture.  Clearly, coupled with a properly designed and cost-effective ICT infrastructure, active consumer participation plays an instrumental role in facilitating the adoption of some of the smart grid technologies and features.}

However, somewhat remarkably, most of the existing research in this area is still based on formal mathematical constructs, such as game theory or classical optimization, which presume that consumers are objective, rational entities that are uninfluenced by real-world behavioral considerations. While such an assumption will hold in a highly-centralized, traditional power system, it will remain an important barrier that has prevented the widespread adoption of the smart grid.

The primary goal of this article is to shed light on the role of consumer participation in the grid, while exposing the role of the Nobel-prize winning framework of prospect theory, in providing a mathematical basis within which to better understand how consumer behavior and realistic smart grid considerations impact the operation and efficiency of smart energy management mechanisms such as DSM, energy trading, and storage management. To this end, we first provide an overview of important energy management services in which consumer participation plays an important role. For each such service, we expose the state of the art and discuss the key assumptions and limitations. Then, we present the basics of prospect theory and discuss the motivation for applying this framework in a smart grid environment. We illustrate the benefits of prospect theory via two simple examples pertaining to energy storage management and DSM. We conclude by providing a future roadmap on how behavioral studies can play an instrumental role in future smart grid designs.

The rest of this paper is organized as follows: Section~\ref{sec:sysmodel} presents an overview on existing energy management literature. In Section~\ref{sec:pt}, we provide a tutorial on the framework of prospect theory and its motivation. Then, in Section~\ref{sec:ptsg}, we discuss two smart grid examples. Finally, Section~\ref{sec:conc} draws some conclusions and future work.

\section{Energy Management in the Smart Grid: A Review}\label{sec:sysmodel}
Owing to the deployment of a smart communication infrastructure and to the presence of new devices, such as storage units, smart meters, and renewable sources, the smart grid presents numerous new opportunities for energy production and distribution that were not possible in a classical grid. For instance, the possibility of deploying smart meters at consumer premises, opens the door for enabling consumers to actively manage their energy. In addition, the ability of a smart meter to communicate in real-time with a power company's control center, provides the latter with various opportunities to actively control and monitor energy usage. Such new capabilities undoubtedly change the way in which energy is generated and distributed to consumers. Clearly, \emph{smart energy management} protocols and mechanisms are needed to exploit the opportunities brought forward by this new smart grid infrastructure.

Deploying efficient energy management mechanisms in future smart grid systems faces many challenges. The first such challenge is to actively utilize smart metering and consumer-based energy management systems to shape the overall grid load over time. Such load shaping is quintessential for an efficient operation of a large-scale smart grid. {Enabling such demand-side management requires both increased automation and active participation by the grid consumers.} Another important challenge is to properly integrate and exploit storage devices in the grid. {For instance, on the one hand, a power company can make use of EVs to store energy reserves so as to regulate the grid operation. In this example, the power company will be submitting an offer for ancillary services to an independent system operator (ISO). On the other hand, a consumer-based storage unit can be used as a means to store or even sell energy back to the power company (rather than an ISO).} Last but not least, an important energy management challenge is to properly decide on how to integrate and utilize consumer-based power sources, such as solar panels, within an operating smart grid system.

In summary, energy management in the smart grid involves the planning and operation of energy-related production and consumption units, particulary when such units are consumer-owned. Next, we summarize some of the main research topics related to energy management in the smart grid.

\subsection{Demand-side Management}
{Demand-side management and demand response programs are arguably the most important form of energy management in the smart grid. DSM can entail a broad range of programs. These programs range from classical direct consumer load control to peak shaving programs and ancillary service provisions. Naturally, each such DSM program has its own challenges. Here, we will summarize some of these programs, while emphasizing the role of consumers. For example, in peak shaving or direct load control programs, DSM schemes typically aim at encouraging consumers to change their energy consumption habits during peak hours.} In particular, such load control DSM schemes aim at providing consumers with incentives to shift their unnecessary grid load to various times during the day, so as to shape the peak hour load on the grid. {For example, a simple DSM scheme can provide monetary benefits to consumers if they use delay-tolerant equipment, such as dish washers or washing machines, during the night, instead of peak hours~\cite{HAM00,DSMDR00,DR02,MGA04,QZ00,DR03,DR04,DSMDR01,DSMDR02,FF05,GT04,FF03,FF04,CP04,CP00,CP01,CP02,CP08,DSMDR03,DSMDR04,DSMDR05,DSMDR06,DSMDR07,DSMDR08,DSMDR09,DSMDR10,DSMDR12,DSMDR13,DSMDR14,DSMDR15,DSMDR16,DSMDR17,DSMDR18}. Even if the individual appliance consumption can be small, the participation of consumers at scale, such as within a neighborhood or city, will significantly impact the smart grid. Moreover, DSM will also extend to other types of consumer-owned devices, such as storage units or renewable energy sources whose consumption and usage might be more significant than standard appliances~\cite{HAM00,DSMDR00,DR02,MGA04,QZ00,DR03,DR04,DSMDR01,DSMDR02,FF05,GT04,FF03,FF04,CP04,CP00,CP01,CP02,CP08,DSMDR03,DSMDR04,DSMDR05,DSMDR06,DSMDR07,DSMDR08,DSMDR09,DSMDR10,DSMDR12,DSMDR13,DSMDR14,DSMDR15,DSMDR16,DSMDR17,DSMDR18}. Last but not least, consumers of DSM need not only be home users, but they can extend to industry and even small, local energy providers that work hand-in-hand with the power company. In such cases, significant gains for both consumers and power companies can be achieved if DSM is properly implemented~\cite{FAHEY7}.}

\subsubsection{Challenges}  The key challenge in DSM is to design realistic incentive mechanisms that can be used by power companies and consumers alike to manage their power consumption over time. The essence of demand-side management revolves around modeling the interactions and decision making processes of various grid players whose goals and actions are largely interdependent. For example, the change in the load of a certain consumer can lead to change in the pricing scheme used by the power company which, in turn, can lead to a change in the behavior of other consumers. This large coupling in the behavior and goals of the grid consumers has led to an abundant literature that applies the mathematical framework of \emph{game theory}~\cite{GTBOOK} to analyze and design efficient DSM schemes.

Game theory is a mathematical framework that enables one to model the decision making processes of a number of players whose objectives are largely interdependent. The merits of a game-theoretic approach for DSM include: 1) ability to capture the heterogeneity of the devices in the grid, 2) effective integration of consumer-based decisions, 3) synergy between game-theoretic designs and the design of incentive mechanisms, and 4) low-complexity learning mechanisms that can characterize the outcome of a game and that can be practically implemented in a real-world smart grid.

\subsubsection{State-of-the-art} There has been {a surge} in research activities related to DSM in recent years~\cite{HAM00,DR02,DSMDR00,MGA04,QZ00,DR03,DR04,DSMDR01,DSMDR02,FF05,GT04,FF03,FF04,CP04,CP00,CP01,CP02,CP08,DSMDR03,DSMDR04,DSMDR05,DSMDR06,DSMDR07,DSMDR08,DSMDR09,DSMDR10,DSMDR12,DSMDR13,DSMDR14,DSMDR15,DSMDR16,DSMDR17,DSMDR18}. As already mentioned, the majority of these works adopts a game-theoretic approach to demand-side management.

One of the earliest works in this area is \cite{HAM00} which presents a DSM model in which the users are able to decide on how to schedule their appliances over a given time horizon. The basic idea is simple: each consumer selects a certain schedule for its appliances, in such a way so as to minimize its overall cost; given a fixed, yet well-designed pricing scheme from the company. Using a game-theoretic model, the authors in \cite{HAM00} characterize the eventual operating system of the grid and show that, under the assumption that users are rational and will act strategically, significant reductions in the overall grid load can be foreseen using such a DSM scheme.

The recent work in \cite{DSMDR00} extends the model of \cite{HAM00} by including the power company as a player in the system. In this regard, following a grid model similar to \cite{HAM00}, the authors enable the power company to strategically decide on its pricing depending on the total power. The objective is to reduce the peak-to-average ratio (PAR) of the load demand. Using numerical simulations, the authors establish the merits of such a DSM scheme and show that noticeable reduction in the PAR can be harnessed via dynamic pricing that adapts to the users' behavior.

Another key contribution on DSM is presented in \cite{DR02}. In this work, the goal is not to reduce peak hour consumption, but rather to match the supply and demand. Depending on whether there is a deficit or excess of energy, the proposed game-theoretic market model incentivizes the consumers to either increase or shed their load so as to match the supply and maintain normal grid operation. Thus, this work highlights an interesting use of DSM for regulating the overall grid operation, rather than just for reducing or shifting load over time.

The work in \cite{MGA04} studies, using a game-theoretic and optimization framework, the ability of consumers to coordinate the way in which they defer their grid load, based on the power company's pricing scheme. In particular, the authors observe that, when DSM protocols leave the smart meters to react independently, in an uncoordinated manner, to pricing fluctuations, new peak hours may be created thus defeating the main purpose of a DSM scheme. To this end, the authors propose a coordination mechanism between a large population of smart grid consumers. In this mechanism, instead of directly reacting to the pricing change, smart meters, acting on behalf of users, aim to adapt the deferment of loads to the changes in the price. One of the key contributions of this work is to consider such a coordination over a large number of consumers. The results, based on realistic, empirical market models from the UK, show that such a coordinated DSM approach can reduce peak hour demand while also reducing carbon emissions.

In \cite{QZ00}, an interesting game-theoretic framework was developed to answer an important question: what is the value of DSM and demand response mechanisms in the smart grid. Essentially, the system is viewed as a noncooperative, hierarchical game between a number of generation companies and the consumers. On the one hand, the generation companies are controlled by the utility operator who can determine their production level. On the other hand, the consumers are aggregated into one collective decision maker, who responds to a pricing signal sent out by the operator, to determine the overall consumption level. Using this model, it has been shown that the use of a demand response mechanism can, in some cases, be more beneficial to generation companies than to consumers. This benefit is largely dependent on how consumers respond to the pricing scheme. Therefore, this work has yet again shown that the way in which consumers behave must be properly modeled if one is to reap the benefits of a technology such as DSM.

Building on those key contributions, a number of equally interesting DSM schemes have emerged more recently~\cite{DR03,DR04,DSMDR01,DSMDR02,FF05,GT04,FF03,FF04,CP04,CP00,CP01,CP02,CP08,DSMDR03,DSMDR04,DSMDR05,DSMDR06,DSMDR07,DSMDR08,DSMDR09,DSMDR10,DSMDR12,DSMDR13,DSMDR14,DSMDR15,DSMDR16,DSMDR17,DSMDR18} expanding on the aforementioned works by developing more advanced models such as those that integrate additional players or other energy efficiency metrics.

\subsubsection{Summary and Remarks} Clearly, existing research has established the technological benefits of DSM. Indeed, most of the existing works such as in~\cite{HAM00,DSMDR00,MGA04,QZ00,DR03,DR04,DSMDR01,DSMDR02,FF05,GT04,FF03,FF04,CP04,CP00,CP01,CP02,CP08,DSMDR03,DSMDR04,DSMDR05,DSMDR06,DSMDR07,DSMDR08,DSMDR09,DSMDR10,DSMDR12,DSMDR13,DSMDR14,DSMDR15,DSMDR16,DSMDR17,DSMDR18}, have shown that under fairly realistic scenarios, DSM can yield significant reduction in peak hour load and can provide an interesting means for regulating the overall operation of the grid. The nexus of these existing works remains a game-theoretic model in which various interactive scenarios between the users and one or more utility providers are modeled. The outcomes include a broad range of pricing mechanisms and load-shifting scheduling algorithms that can be implemented to optimize various energy efficiency metrics, such as peak hour load, PAR, and load regulation metrics. Undoubtedly, DSM and related ideas are likely to become an important component of the smart grid.

Alas, despite this established gains of DSM, the real-world implementation of such programs (and related ideas) has remained below expectations~\cite{FAHEY,FAHEY2,FAHEY3,FAHEY4,FAHEY5,FAHEY6,FAHEY7}. One of the underlying reasons is that, in real life, consumers are not behaving the way they are supposed to, as assumed by many existing mathematical models. In this regard, most of these existing models rely on the assumption that players are \emph{rational} and will act objectively when faced with a DSM decision. In other words, these models presume that, in the real-world, consumer behavior will follow strictly objective measures of benefits and losses, when deciding on whether or not to subscribe to a DSM scheme or when choosing on how to schedule appliance usage. However, realistically, consumers may deviate from the rational behavior to various factors. For example, on a cold winter day, consumers may be reluctant to shift their heating consumption to a later time of the day, even if such a shift can be beneficial to the grid or can bring some economic benefit to the consumers.

Clearly, within the context of DSM, there is an urgent need to capture such ``behavioral'' factors when designing the demand response mechanisms of the future. Without a careful accounting for the behavioral side, the real-world adoption of DSM mechanisms will not live up to the expectations.

\subsection{Integration of Storage Units and Consumer-Owned Renewable Sources}
Beyond DSM and demand response mechanisms, energy management in the smart grid must account for the presence of a variety of new devices that are expected to be deployed in the near future. Such devices include EVs, storage units, and renewable energy sources. While renewable energy sources may be owned by energy providers or consumers, the majority of EVs and storage units are expected to be consumer-owned. The presence of such new components in the power grid presents an interesting opportunity for deploying evolved energy trading mechanisms~\cite{ET00,ET01,ET02,ET03,ET04,ET05,ET06,WI03}.

In particular, the ability of consumers to store energy or possibly feed energy back into the grid, via either their storage units or their owned renewable sources, will pave the way towards a large-scale exchange of energy within the grid. For example, on the one hand, consumers with a surplus of energy may decide to send this energy back into the grid, to improve grid regulation and reap some possible monetary benefits. On the other hand, the power company may utilize EVs or other storage units as a means to store energy reserves or to regulate the grid frequency~\cite{STOR00,STOR01,STOR02,EW07,EW08,EW02,EW05,EW06}. Indeed, if properly managed, the charging and discharging behavior of EVs and storage units can yield significant technological and economic benefits for power companies and consumers alike~\cite{MGA,MGA04,GT06,HAM01,WI03,REN00,STOR00,STOR01,STOR02,STOR03,STOR04,STOR05,STOR06,STOR07,EW01,EW07,EW08,EW02,EW05,EW06,ET00,ET01,ET02,ET03,ET04,ET05,ET06,ET07,ET08,ET09}.

In a nutshell, the effective integration and exploitation of storage units and consumer-owned renewable sources will be an essential property of the future smart grid. How to efficiently exploit such devices to improve the delivery, production, and management of smart grid energy is thus an important problem that must be addressed.

\subsubsection{Challenges} The challenges of integrating energy storage and renewable sources are numerous. From a power system point of view, the intermittent nature of renewable sources will require fundamentally new ways to operate energy production and generation in the grid. How to develop stochastic optimization algorithms that can adapt to this intermittency is thus a key challenge. In addition, the foreseen large-scale deployment of EVs will present an unprecedented increase in the load on the grid. Here, effective DSM mechanisms, as those discussed in the previous section, which can manage the EVs load will be needed. Nonetheless, storage units and EVs also provide an opportunity for the power grid to store any excess or mismatch in the generation and demand, so as to regulate the overall grid operation.

More relevant to this article are the challenges pertaining to the use of storage units and consumer-owned renewables within energy trading markets. In particular, it is foreseen that local markets in which consumers may directly exchange energy with one another or with the grid can be set up in the future smart grid. Such markets are enabled by the presence of storage units, EVs, and consumer-owned generation sources. Important challenges here include: 1) devising economic mechanisms that incentivize consumers, power companies, and energy providers to setup such markets, 2) analyzing the impact of such localized markets on grid operation, and 3) integrating such energy trading within existing DSM mechanisms.

\subsubsection{State-of-the-art}
Integrating storage units and renewable energy sources has been a topic of significant interest to the smart grid community in recent years~\cite{MGA,MGA04,GT06,HAM01,WI03,REN00,STOR00,STOR01,STOR02,STOR03,STOR04,STOR05,STOR06,STOR07,EW01,EW07,EW08,EW02,EW05,EW06,ET00,ET01,ET02,ET03,ET04,ET05,ET06,ET07,ET08,ET09,MIC00,MIC01,MIC02,MIC03,MIC04,MIC05,MIC06,MIC07,MIC08,HP05}. Beyond the works that focus primarily on the power system operation side~\cite{MIC00,MIC01,MIC02,MIC03,MIC04,MIC05,MIC06,MIC07,MIC08,HP05}, there has been a number of interesting works that investigate the usage of storage, EVs, and renewables, to shape the overall grid load and to establish energy trading markets~\cite{MGA,MGA04,GT06,HAM01,WI03,REN00,STOR00,STOR01,STOR02,STOR03,STOR04,STOR05,STOR06,STOR07,EW01,EW07,EW08,EW02,EW05,EW06,ET00,ET01,ET02,ET03,ET04,ET05,ET06,ET07,ET08,ET09}.

The earliest work in this area is in \cite{MGA} in which a game-theoretic framework is developed to analyze how consumers equipped with storage unit can smartly decide on when to buy or store energy, in a local smart grid area. The presented scheme is essentially a modified DSM protocol which explicitly factors in the presence of storage units. The market price is assumed to be pre-determined using an auction mechanism and, thus, the work does not account for dynamic pricing. Simulation results presented in \cite{MGA04} show that, based on empirical data from the UK market, the use of storage at consumer premises along with game-theoretic DSM protocol can help in reducing the peak demand which also leads to reduced costs and carbon emissions. The results also analyze the benefit of storage and how it impacts the social welfare of the system thus highlighting the possible practical impact of storage unit integration.

One of the most interesting works that follows in this direction is presented in \cite{GT06}. In this contribution, the authors study a DSM-like scheme in which users can be endowed with storage units and renewable sources. In contrast to traditional DSM schemes in which users only decide whether or not to purchase energy from the grid, the model in \cite{GT06} enables the users to decide on whether to purchase, produce, or store energy in their batteries. By expanding the decision space of the users, it is shown, using a game-theoretic approach, that a smart exploration of the storage and energy production options can reduce the overall aggregate load on the grid while also providing monetary savings to end-users, under the assumption of rational decision making. Such a study thus motivates the penetration of consumer-owned storage and energy production units.

The effective integration of EVs into a smart grid system is studied in \cite{HAM01} using a game-theoretic framework that models the interactions between the grid operator and the EVs. The primary goal is to analyze how EVs can provide ancillary services to the grid, once a proper market model between EVs and the grid is setup. The basic idea is to use a smart pricing policy to exploit the EVs for regulating the grid frequency. The idea is to study how EVs (and their owners) can decide on whether to charge, discharge, or remain idle, in a way to optimize the grid frequency regulation while benefiting both consumers and the grid operator. On the one hand, using such a scheme consumers can obtain additional income while, on the other hand, the grid can achieve the required frequency regulation command signal.

The impact of energy trading between owners of EVs is further analyzed in our earlier work in \cite{WI03}. In this work, a local market in which EV owners can decide on whether or not to sell a portion of their stored energy to the smart grid is studied. Using an auction and a game-theoretic model, we have shown that, when EV owners act strategically, they are able to reap significant benefits from selling their surplus of stored energy to potential buyers in the smart grid. These benefits are reflected in terms of revenues that can be viewed as either direct monetary gains or as coupons or other offers provided from the grid owner to active participants.

The impact of distributed renewable energy sources on local energy trading markets is analyzed in \cite{REN00}. The authors essentially propose the use of an aggregator of distributed energy resources allowing the smart grid to engage in an open energy exchange market. The developed mechanism tightly integrates classical DSM ideas with the use of an active management scheme at the end-users side to allow a better utilization of the renewable energy sources. Overall, the results show that a smart exploration of possibly consumer-owned energy sources can lead to a sustainable source of energy and a reduction in the consumers' energy consumption cost.

Beyond the aforementioned contributions, the exploitation of storage and renewable sources at consumer premises has been studied in a broad range of literature~\cite{STOR00,STOR01,STOR02,STOR03,STOR04,STOR05,STOR06,STOR07,EW01,EW07,EW08,EW02,EW05,EW06,ET00,ET01,ET02,ET03,ET04,ET05,ET06,ET07,ET08,ET09}. These works mainly develop variants of the discussed energy trading mechanisms and establish clearly that the use of mathematical optimization frameworks such as game theory to manage the way in which storage devices, EVs, and consumer-owned energy sources are integrated in the smart grid can bring in substantial technological, economic, and environmental benefits to the smart grid players.

\subsubsection{Summary and Remarks}
The use of energy trading between consumer-owned devices will indisputably be an important feature of the smart grid. As demonstrated in the abundant literature, the associated gains, both from a technical (energy reduction, sustainable generation) and  an economic (reduced costs on consumers and providers) point of view, are substantial. Yet, despite these established results, beyond some small deployments of EVs and renewable energy sources in Europe and some areas of the United States~\cite{MOMA00,MOMA01,NREL,NREL1,NREL2,NREL3,NREL4,NREL5}, the large-scale introduction of such consumer-owned devices remains modest.

Similar to the DSM case, one of the primary limitations of existing models is that they often do not explicitly factor in realistic risk considerations of both consumers and power companies. For instance, even though an open, energy trading market can yield economic and technological benefits, power companies may remain risk averse and continue to rely on traditional, largely controlled markets. Similarly, despite the prospective economic savings and environmental benefits of owning renewable sources or EVs, consumers may still be reluctant to change from their current, effective technologies.

Therefore, when analyzing energy trading, one must explicitly factor in such risk considerations and their impact on the overall operation of smart energy management mechanisms.

\section{Prospect Theory for the Smart Grid}\label{sec:pt}
\subsection{Introduction and Motivation}\label{sec:pt2}
As demonstrated in Section~\ref{sec:sysmodel}, the existing literature on smart energy management in the smart grid has established significant technological, economic, and environmental benefits for features such as DSM and energy trading. Yet, the real-world deployment of such mechanisms remains largely below expectations. One of the hurdles facing the real-world implementation of the developed DSM and energy management mechanisms, is the lack of a mathematical and empirical framework that can capture the realistic behavioral patterns of consumers and power companies.

Indeed, most existing works~\cite{HAM00,DR02,DSMDR00,MGA04,QZ00,DR03,DR04,DSMDR01,DSMDR02,FF05,GT04,FF03,FF04,CP04,CP00,CP01,CP02,CP08,DSMDR03,DSMDR04,DSMDR05,DSMDR06,DSMDR07,DSMDR08,DSMDR09,DSMDR10,DSMDR12,DSMDR13,DSMDR14,DSMDR15,DSMDR16,DSMDR17,DSMDR18,MGA,GT06,HAM01,WI03,REN00,STOR00,STOR01,STOR02,STOR03,STOR04,STOR05,STOR06,STOR07,EW01,EW07,EW08,EW02,EW05,EW06,ET00,ET01,ET02,ET03,ET04,ET05,ET06,ET07,ET08,ET09,MIC00,MIC01,MIC02,MIC03,MIC04,MIC05,MIC06,MIC07,MIC08,HP05} still assume that consumers and power companies are rational and will abide by the objective decision making rules that are derived via frameworks such as game theory or optimization. However, in practice, empirical studies have shown that, in uncertain and risky situations, human players may not act in accordance with the rational behavior established by decision making frameworks such as game theory~\cite{PT00,PT01,PT02,PT03,PT04}. Given that most foreseen smart grid features are consumer-centric, the ``human'' factor will undoubtedly play an instrumental role in the success or failure of advanced smart grid features. Thus, uncertainty and risk factors must be properly modeled in any DSM or energy management scheme. Examples of risk in the smart grid include the continuous reliance of operators on traditional markets and the interdependence of the decisions between consumers. In terms of uncertainty, when dealing with an energy management scheme, consumers are faced with uncertain outcomes due to a lack of transparency in explaining the rules of dynamic pricing or due to the presence of stochastic elements such as stochastic generation or uncertain presence or absence of EVs and energy storage units.

Thus, expediting the smart grid adoption requires new approaches for analyzing the often irrational and non-conforming nature of the energy management decisions of human players under  such risk and uncertainty. Such decision-making factors
that deviate from the objective, rational behavior assumed in existing works~\cite{HAM00,DSMDR00,QZ00,DR03,DR04,DSMDR01,DSMDR02,FF05,GT04,FF03,FF04,CP04,CP00,CP01,CP02,CP08,DSMDR03,DSMDR04,DSMDR05,DSMDR06,DSMDR07,DSMDR08,DSMDR09,DSMDR10,DSMDR12,DSMDR13,DSMDR14,DSMDR15,DSMDR16,DSMDR17,DSMDR18,MGA,MGA04,GT06,HAM01,WI03,REN00,STOR00,STOR01,STOR02,STOR03,STOR04,STOR05,STOR06,STOR07,EW01,EW07,EW08,EW02,EW05,EW06,ET00,ET01,ET02,ET03,ET04,ET05,ET06,ET07,ET08,ET09,MIC00,MIC01,MIC02,MIC03,MIC04,MIC05,MIC06,MIC07,MIC08,HP05} can be analyzed
by the Nobel-prize winning framework prospect theory (PT)~\cite{PT00,PT01,PT02,PT05}. Originally conceived for modeling decisions during monetary transactions such as lottery outcomes, PT has
made its way into many applications~\cite{PT00,PT01,PT02,PT03,PT04,PT05,NM00,NM01}, due to the universal applicability of its concepts.  In essence, PT provides one with mathematical tools to understand how decision making, in real life, can deviate from the tenets of expected utility theory (EUT), a conventional game-theoretic notion which is guided strictly by objective notions of gains and losses, player rationality, conformity to pre-determined decision making rules that are unaffected by real-life perceptions of benefits and risk.

\textbf{Illustrative Example:} Essentially, PT notions have been developed to understand how consumers, when faced with uncertainty of outcome and risky decisions, will behave in real-life. Suppose that an efficient energy management system is constructed for individual home owners to both buy and sell power on the grid and a dynamic pricing DSM mechanism is available to shift consumption to non-peak periods. Furthermore, suppose that it has been proven that under PT as well as conventional game theory, stable prices can be found, so that the smart grid could ultimately result in more efficient power consumption. Under \emph{rational analysis}, one might believe when these conditions were satisfied, offering the opportunity to buy and sell power to the public would result in widespread participation and an optimal pricing equilibrium would soon be reached. However, an important implication of PT is that these conditions are insufficient to guarantee such a beneficial result.

One important implication of PT is that the preferred choice between a pair of uncertain alternatives is not only determined by the values of the two alternatives but also by how the choice is stated. Consider the following example, which is unnatural only in that the alternatives are designed to have equal value, so that a preference is clearly determined by the statement of the choice. A power company wishes to entice its consumers to abandon buying power at a fixed rate and instead join a system where they buy and sell power at variable rates. Here are two ways the alternatives may be presented in a letter to a consumer:
\begin{itemize}
\item \emph{The Gain Scenario:} Your average monthly utility bill is now \$450 a month. Under our new smart system, your bill will show a debit of \$500 a month.  In addition you may choose between:\\

a)	A $50\%$ chance of a credit of \$100 if you join the smart DSM scheme, or\\
b)	A $100\%$ chance of a credit of \$50 that will keep your bill the same.
\item \emph{The Loss Scenario:}
Your average monthly utility bill is now \$450 a month. Under our new smart system, your bill will show a credit of \$400 a month. In addition, you may choose between:\\
c)	A $50\%$ chance of a bill for \$100 if you join the smart DSM scheme, or\\
d)	A $100\%$ chance of a bill for \$50 that will keep your bill the same.
\end{itemize}

In fact, the Gain and Loss scenarios describe the identical alternatives in different words. Alternatives a) and c) are identical and alternatives b) and d) are identical. Nevertheless, based on theoretical and empirical foundations, PT predicts that more people will prefer alternative b) to alternative a) because a certain gain is preferred to a $50\%$ chance at a double-gain but will also prefer alternative c) to alternative d) because a $50\%$ chance of a loss is preferred to a certain, albeit smaller, loss. This prediction has been confirmed in~\cite{PT00} and \cite{CPT00}.

The point of this example is not just that the level of participation in smart grid services depends on how it is presented to the public. The point is that important behavioral factors outside of the technical specifications of the smart grid will determine the choices of participants and giving them the opportunity to perform optimally does not guarantee that they will. In other words, people cannot be counted on to always choose optimally among alternatives if merely stating the alternatives differently influences their choices. This holds true even if such alternatives, as discussed in Section~\ref{sec:sysmodel}, have immense technological and environmental benefits. Indeed, in \cite{KAH00}, Kahneman suggests that people behave non-optimally when buying and selling stocks, selling rising stocks too soon to lock in gains and hanging on to losing stocks too long to resist locking in a loss. If people behave non-optimally in the purchase and sale of securities, the default assumption is that they will perform in the same non-optimal manner in the purchase and sale of power and commodities, especially when people are already familiar with the incumbent pricing and energy management mechanism.

The obvious solution to the problem of human behavior is to use prospect-theoretic notions to refine existing game-theoretic mechanisms and guide the way in which optimal strategic decisions are derived as well as to improve the presentation of information to buyers and sellers in the grid to encourage optimal behavior in DSM and energy trading. To provide further insights on the mathematical machinery underlying PT, in the next subsection, we provide an introduction to the basics of the framework.

\subsection{Basics of Prospect Theory}
Prospect theory encompasses a broad range of techniques and tools to account for realistic consumer behavior during decision making processes~\cite{PT01,PT02,PT03,PT04,PT05,CPT00,KAH00}. The basic underlying idea is that decision makers, in real-life, will have subjective perceptions of losses, gains, and their competitive environment. For example, instead of viewing each others' actions (e.g., load shifting schedules) objectively as in classical game theory, players could have different subjective assessments about each others behavior, which, in turn can lead to unexpected, irrational decisions. For example, in DSM, even though rational behavior dictates that consumers follow the load shifting mechanisms of the power company, some consumers may turn on certain appliances at unexpected times, since they are unsure on whether participation level is high enough to obtain economic benefits which will hinder the performance of the DSM scheme. The large spread of such unconventional actions can thus be disruptive to any energy management scheme. In such situations, PT provides solid analytical tools that directly address how these choices are framed and evaluated, given the subjective observation of players in the decision-making process.

\subsubsection{Subjective Actions -- The Weighting Effect}  The first important PT notion is the so-called \emph{weighting effect}. In particular, in PT~\cite{PT00,PT01,PT02,PT03,PT04,PT05}, it is observed that in real-life decision-making, people tend to subjectively weight uncertain
outcomes. {In particular, in energy management mechanisms, the frequency with which a consumer chooses a certain strategy, say a certain schedule of appliances or a certain storage pattern, depends on how other consumers make their own choices. The dependence stems from many factors. For example, in dynamic pricing schemes, the actual price announced by a power company depends on the entire load of the consumers. Therefore, the decision of a consumer, will subsequently depend on the decision of others. Indeed, when faced with a given smart grid scenario, consumers may act differently over time due to the interdependence of their actions and its unpredictability, and, thus, a probabilistic model for decision making is suitable to capture this uncertainty. Such uncertainty can stem not only from the individual decisions of consumers but also from other smart grid factors (e.g., uncertainty of renewable energy).}

In classical game-theoretic smart grid schemes~\cite{HAM00,DSMDR00,QZ00,DR03,DR04,DSMDR01,DSMDR02,FF05,GT04,FF03,FF04,CP04,CP00,CP01,CP02,CP08,DSMDR03,DSMDR04,DSMDR05,DSMDR06,DSMDR07,DSMDR08,DSMDR09,DSMDR10,DSMDR12,DSMDR13,DSMDR14,DSMDR15,DSMDR16,DSMDR17,DSMDR18,MGA,MGA04,GT06,HAM01,WI03,REN00,STOR00,STOR01,STOR02,STOR03,STOR04,STOR05,STOR06,STOR07,EW01,EW07,EW08,EW02,EW05,EW06,ET00,ET01,ET02,ET03,ET04,ET05,ET06,ET07,ET08,ET09,MIC00,MIC01,MIC02,MIC03,MIC04,MIC05,MIC06,MIC07,MIC08,HP05}, consumer interdependence is captured via the notions of expected utility theory in which a consumer computes an expected value of its achieved gains or losses, under the observation of an objective probability of choice by other consumers. In contrast, using the weighting effect, PT allows one to capture each consumer's subjective evaluation
on the probabilistic strategies of its opponents. Thus, under PT, instead of objectively observing the information given by the other players and computing a classical expected value for the utility, each consumer perceives a weighted version of its observation on the other actions. The weighting is used to express a ``distorted'' view that a given consumer or player can have on the actions of others. {PT studies have shown that most people overweight low probability outcomes and underweight high probability outcomes.}

\begin{figure}[!t]
  \begin{center}
   \vspace{-0.2cm}
    \includegraphics[width=8cm]{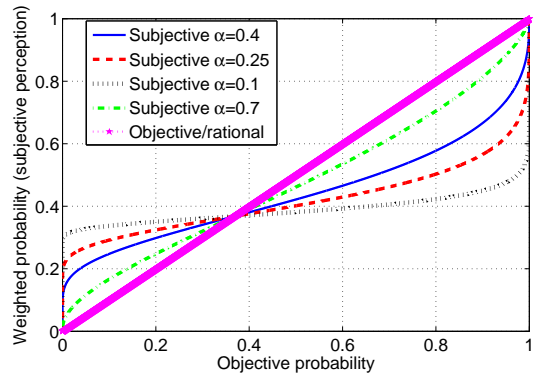}
    \vspace{-0.3cm}
    \caption{\label{fig:ill} Illustration of the prospect-theoretic weighting effect: how objective probabilities are viewed subjectively by human participants. The parameter $\alpha$ determines how far the behavior is from the fully rational case.}
  \end{center}\vspace{-1cm}
\end{figure}

To illustrate the weighting effect, in Fig.~\ref{fig:ill}, we show an example of a weighting function, known as the Prelec function~\cite{PRELEC}, which maps an objective utility into a subjective utility. The mapping is controlled by a parameter $\alpha$ which quantifies the level of subjectivity in the observation. For $\alpha=1$, we have the fully rational case, while for $\alpha$ close to $0$ we get the fully irrational case. Within a smart grid setting, such a weighting can have a cascading effect on the way in which an energy management scheme works. For example, in a DSM context, a highly irrational consumer will have a largely distorted view on how other consumers behave. In turn, this consumer will become more risk averse or more risk seeking, depending on how the opponents actions impact the dynamic pricing mechanism. As a result, this consumer will not follow the actions recommended by classical, rational mechanisms, but, instead will take an unexpected action which, in turn, will yield unexpected DSM results.

Indeed, how to model such a weighting effect and how to integrate it into realistic energy management mechanisms is an important topic for research. In addition, how to design weighting functions that are tailored towards the smart grid and that can work in realistic power system setting is a key challenge. In Section~\ref{sec:ptsg}, we will discuss with specific examples how weighting modifies the results of energy trading and management protocols.

\subsubsection{Subjective Perceptions of Utility Functions -- The Framing Effect} Another important idea brought forward by PT is the notion of \emph{utility framing}. In engineering designs, one often defines mathematically rigorous objective (utility) functions that are used to optimize a certain metric of the system. For example, when dealing with an optimal energy generation problem, a smart grid system must find the maximum energy output that can meet or match the demand. In such a case, it is sound to assume that the function that must be optimized is based on an objective metric, the energy in this case.

However, when dealing with smart energy management mechanisms having human players, the idea of an objective metric for evaluating utility functions might not be a reasonable assumption. For instance, each individual has a different perception on the economic gains from a certain DSM scheme. For example, a saving of \$10 per month may not seem like a significant gain for a relatively wealthy consumer. Instead, a poor consumer might view this amount as a highly significant reduction. Clearly, the objective measure of \$10, can be viewed differently by different consumers.

\begin{figure}[!t]
  \begin{center}
   \vspace{-0.2cm}
    \includegraphics[width=8cm]{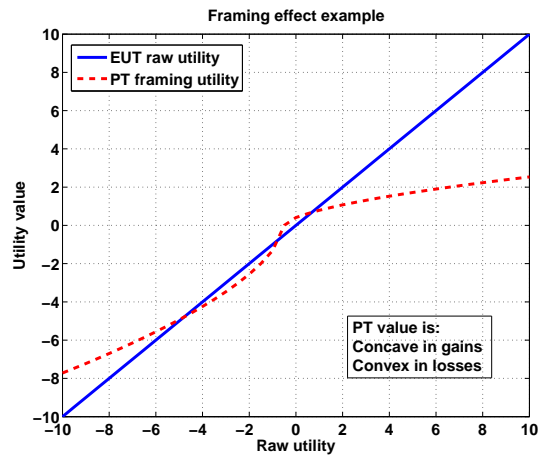}
    \vspace{-0.3cm}
    \caption{\label{fig:fra} Illustration of the prospect-theoretic framing effect: how objective utilities are viewed subjectively by human participants. The utility function value changes depending on a certain reference point that highlights the individual perceptions of gains and losses.}
  \end{center}\vspace{-0.3cm}
\end{figure}

In PT, such subjective perceptions of utility functions are captured via the idea of framing or reference points. In essence, each individual frames its gains or losses with respect to a possibly different reference point. Back to the aforementioned example, the wealthy consumer will frame the \$10 with respect to its initial wealth which could be close to millions and, thus, this consumer views the \$10 as insignificant. In contrast, the poor consumer might have a wealth close to $0$ and, thus, when framing the \$10 with respect to this reference point, the gains are viewed as significant. One popular way to capture such framing effects is by observing that losses loom
larger than gains, and, thus,  PT provides one transformation that maps objective utility functions into so-called subjective value functions -– concave in gains, convex in losses –- over the possible outcomes. These gains and losses are measured
with respect to a reference point that need not be 0 and that may be different between players. An illustrative example is shown in Fig.~\ref{fig:fra} for one typical PT value function from~\cite{PT00} assuming a zero reference point for gains/losses.

Naturally, as consumers change the way in which they compute their utilities, their overall decision making processes will deviate from the conventional, rational thought. Indeed, when applying PT ideas to game-theoretic settings such as in \cite{PT05}, it is shown that the objective results do not hold. For example, in some cases, it is shown that the choice of a reference point can impact whether or not a certain game has an equilibrium solution or not. Clearly, when one decision maker changes the way in which it evaluates its objective function, the overall operation of any optimization mechanisms will be significantly affected.

In the smart grid, we can envision many situations in which to incorporate the framing effects. These situations need not be purely economical.
{For example, during winter, consumers may perceive less prospective gain from turning off high-capacity loads (such as heaters) at night than during day time.} How this ``frame of reference'' transforms the utility will fundamentally change the outcome of an energy management mechanism that is based on classical objective notions. Moreover, in the smart grid, such reference points and framing effects may change over time, space, and even demographics. Clearly, properly designing and developing framing notions in smart grid DSMs is an important direction that must be investigated to better understand its impact on energy management and trading mechanisms.

Having defined the two key effects of PT, in the next section, we discuss, in detail, two energy management scenarios to highlight, as an example, the impact of weighting on smart grid protocols.

\section{Prospect-Theoretic Smart Grid Applications}\label{sec:ptsg}
\subsection{Example 1: Charging and Discharging of Consumer-Owned Energy Storage}
To show the impact of prospect-theoretic considerations in smart grid design, we first study a model in which consumers are equipped with storage units and must decide on how to manage the charging and discharging of their storage units, depending on the network state and the pricing incentives. This model is based on our recent work in \cite{WSM00}.
\begin{figure}[!t]
  \begin{center}
   \vspace{-0.2cm}
    \includegraphics[width=8cm]{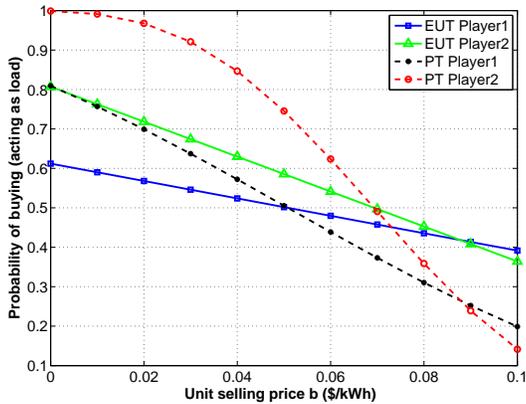}
    \vspace{-0.3cm}
    \caption{\label{fig1} Impact of the weighting effect on consumer behavior in a two-player charginge/discharging setting.}
  \end{center}\vspace{-0.4cm}
\end{figure}

In particular, we consider a grid consisting of multiple consumers who own storage units. For illustration purposes, we assume the case in which only two consumers are ``active participants'' while all other consumers constitute a passive load on the grid. Each consumer has a storage unit which holds a certain initial amount of energy stored. The power company offers these active consumers the option to either charge their storage unit and, thus, act as a load on the grid, or, instead to actively feed back and sell energy to the grid. Note that, any action taken by either of the two consumers affects both the power system as it impacts the overall needed generation as well as the prices set by the utility company. The choices of both consumers are also coupled, since the choice of acting as load or source, will impact the overall generation and distribution of energy in the grid.

In this setting, we assume that the consumers need to make a choice between charging or discharging while optimizing a utility function that captures two properties: a) the economic and technical benefits of storing or selling energy and b) the power system regulatory penalties. Indeed, although the power company allows the two active consumers to individually manage their storage units, it still requires the generation to remain within desired operating conditions which are measured based on an initial point.

We formulate and investigate this setting using a PT-based, classical noncooperative game and we study the equilibrium solution of the game. The equilibrium is essentially a point of the system in which neither of the two consumers can improve its utility by changing the frequency with which it chooses to charge or discharge its storage unit. We analyze the results under both classical EUT and PT considerations. For PT, we first consider the weighting effect: each consumer views a subjective observation on the charging/discharging behavior of its opponent in accordance with a Prelec weighting function such as in Fig.~\ref{fig:ill}.

We use a numerical example to show the impact of PT considerations on the operation of the system. We consider a standard $4$-bus power system with $2$ active consumers.  The loads and surpluses of active consumer $1$ are, respectively, $20$~kWh and $10$~kWh, while those of consumer $2$ are, respectively, $15$~kWh and $5$~kWh. Fig.~\ref{fig1} shows the impact of the unit selling price $b$ that is used by the two consumers when discharging energy to the grid. This price is assumed to be equal for both consumers. Clearly, as $b$ increases, both consumers have more incentive to sell than to buy, as the gains start outweighing the regulation penalty. More interestingly, Fig.~\ref{fig1} shows that, for both consumers, PT behavior significantly differs from the rational EUT case. For consumer $2$, below $0.07\$$ per kWh, the probability of buying energy at the PT equilibrium is much higher than under EUT. This implies that for low gains and high risks, the consumer follows a conservative, \emph{risk-averse} strategy under PT and is less interested in reaping the gains of selling energy compared to EUT. However, as the unit selling price crosses the threshold, the probability that consumer $2$ acts as load under PT becomes much smaller than EUT. Thus, once the  selling benefits are significant (risks decrease), PT predicts that consumer $2$ will start selling more aggressively than in EUT. Analogous behavior is seen for consumer $1$ with threshold $0.045\$$ per kWh.

\begin{figure}[!t]
  \begin{center}
   \vspace{-0.2cm}
    \includegraphics[width=8cm]{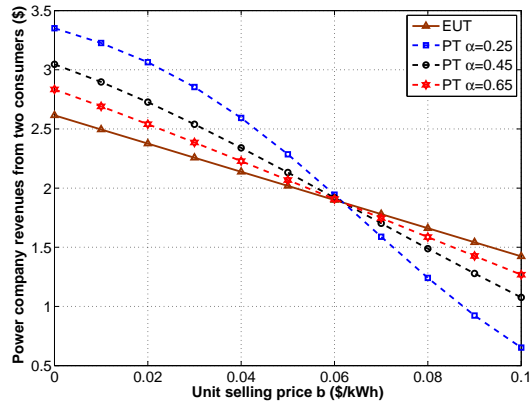}
    \vspace{-0.3cm}
    \caption{\label{fig2} Impact of the weighting effect on the revenues of the power company.}
  \end{center}\vspace{-0.7cm}
\end{figure}

Fig.~\ref{fig2} shows the total power company revenues, collected from the two consumers when they charge their storage unit, as the
 unit selling price increases. Fig.~\ref{fig2} shows that, as $b$ increases, the total revenues decrease, as the consumers begin to sell more and buy less. {Note that, here, the power company's revenues pertain to only those revenues that are collected from the two consumers. This does not include any additional sources of revenues that the power company might collect (e.g., taking a percentage on the profits of the consumers).} Clearly, deviations from  EUT can have major impacts on energy management in a smart grid setting. Consider the case in which the Prelec rationality factor is set to $\alpha=0.25$. When $b$ is below $0.06\$$ per kWh, under PT, the total revenue is much higher than predicted by EUT. In contrast, if consumers set prices greater than $0.06\$$ per kWh, PT predicts that the revenues will be much smaller than in EUT. It is thus more beneficial for the company to regulate the consumers' unit selling price to be below $0.06\$$ per kWh. Fig.~\ref{fig2} also shows that when the company adopts EUT to regulate the consumers' selling price, it can lose revenues due to real-life consumer behavior. Fig.~\ref{fig2} also shows that, as $\alpha$ increases, the consumers behave more in line with EUT. However, even for a relatively high value, $\alpha=0.65$, the company revenues resulting from PT still yield non-negligible deviations from EUT.

\begin{figure}[!t]
  \begin{center}
   \vspace{-0.2cm}
    \includegraphics[width=8cm]{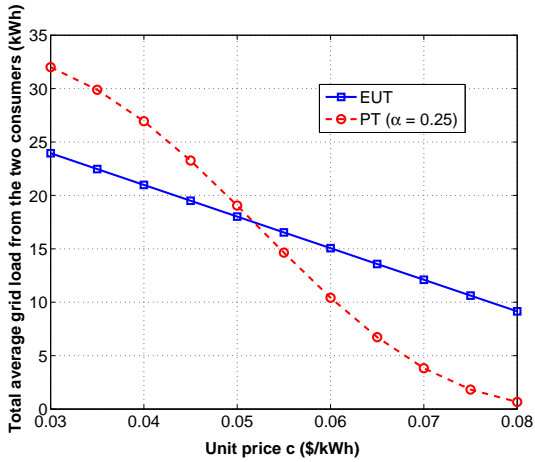}
    \vspace{-0.3cm}
    \caption{\label{fig3} Expected grid load when the consumers actively participate with their storage devices under rational EUT and irrational PT behavior.}
  \end{center}\vspace{-0.7cm}
\end{figure}

We further analyze how consumer behavior impacts grid operation by showing the average expected load on the grid in Fig.~\ref{fig3}, as the company varies its minimum price \footnote{{This pertains to the rates that the power company will use to directly charge its consumers.}}. Fig.~\ref{fig3} shows that the expected load on the system will significantly change between PT and EUT. For PT, when the unit price is small, consumers are less interested in selling their stored energy. However, as the price crosses a threshold, the consumers will start selling more aggressively, rendering the average load smaller than expected. Fig.~\ref{fig3} provides \emph{guidelines for realistic DSM with storage}. For example, assume the company wants to increase its price to drive consumers to sell more and reduce their load to about $10$ kWh. Based on classical EUT-based schemes, the company has to increase the price to $0.078\$$ per kWh. In real life, because consumers behave subjectively under risk, the power company can increase its unit price to only $0.06\$$ per kWh and obtain the desired load reduction. Also, if the power company wants to reduce its price to sustain a load of $23$ kWh from the two consumers, based on EUT, it must offer a price of $0.035\$$ per kWh. In contrast, PT shows that $0.047\$$ per kWh will achieve the same impact yet yield more profits.

\begin{figure}[!t]
 \begin{center}%244
 \vspace{-0.2cm}
  \includegraphics[width=8cm]{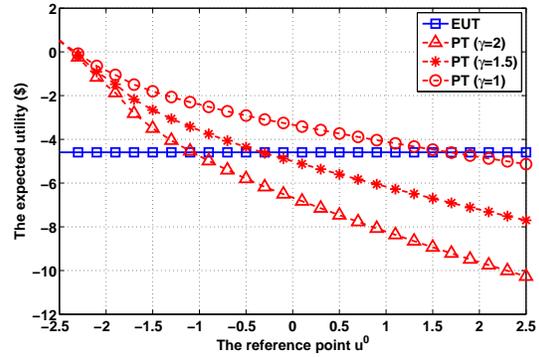}
 \vspace{-0.4cm}
   \caption{\label{fig:UVref} The total utility under both EUT and PT as the reference point $\boldsymbol{u}^0$ varies.}
\end{center}\vspace{-0.9cm}
\end{figure}
{Next, we consider the same example in the presence of framing effects, based on our work in \cite{YW00}. Here, we assume that both consumers frame their utility with respect to a given reference point that reflects how these consumers evaluate the economic gains or losses from charging or discharging their storage unit. For incorporating framing, we adopted the classical model of \cite{PT05} in which, under framing, the utility function becomes concave in gains and convex in losses, as losses loom larger than gains. To assess the impact of framing, in simulations, the reference point is chosen to coincide with the case in which consumers discharge/sell energy at the same price that is announced by the power company. In Fig.~\ref{fig:UVref}, we show the total expected utility (sum for both consumers) under both EUT and PT, as the reference point varies. In this figure, we choose the same reference point for both consumers and we use typical values for $\gamma$ which is a parameter that represents the loss aversion, i.e., how a consumer values its losses versus its gains. First, we can see that, the expected PT utility will decrease when the reference point value increases. In essence, the reference point is subtracted from the EUT utility to determine the exact values of gains and losses. For a high reference point (i.e., electricity price), PT consumers will value their stored energy more than in cases in which the reference point is smaller. Thus, using a same selling/discharging price under EUT and PT, the payoff obtained by consumers under PT is smaller than under EUT, due to the fact that, in PT, the reference point reduces the gains from selling energy. Second, this figure also shows that, the framing aversion parameter, i.e., $\gamma$, would have different impacts on the PT utility. In particular, when $\gamma$ increases from $1$ to $2$, the losses viewed by PT consumers will increase. Thus, with an increasing $\gamma$, a PT consumer will start valuing its gains less than in the EUT case, which leads to increasing its conservative, charging strategy. Additional results on framing can be found in \cite{YW00}.}

In summary, ignoring consumer behavior in storage-based energy management can lead to unexpected results as shown here for a basic setting. These results can be undesirable from both an economic and technological perspective. Therefore, building on the presented model, one can design more elaborate and realistic storage management mechanisms that account for PT-based notions of subjective perceptions. In addition, the power company can utilize these results to properly shape its pricing schemes.

\subsection{Example 2: Demand-side Management under Prospect-Theoretic Considerations}
Another important application for PT is classical demand-side management models. Here, we consider a grid in which consumers are given the opportunity to decide on whether or not to participate in DSM. The DSM scheme considered is one in which the participating consumer would shift its load over time, in order to reduce the overall peak hour load. The actual DSM process is in line with classical game-theoretic settings such as those in \cite{HAM00}.

However, in our model, it is assumed that consumers have also a choice of not participating in the DSM at all. In addition, consumers can choose the time starting which they will begin their participation. The decisions of the consumers are driven by the goal of minimizing the overall electricity bill while maintaining their desired load to operate their required appliances\footnote{The reader interested in the mathematical formulation is referred to our work in \cite{WSM000}.}. One important feature of the considered model is that every load shift by a given consumer will automatically impact the way in which prices are set by the power company. Thus, this interdependence in decision making will naturally warrant a game-theoretic approach to modeling the decision making.

In essence, we have a model in which every consumer can decide on the time at which it starts to participate in DSM. Alternatively, the consumer may decide not to participate at all. We can then analyze the frequency with which a consumer will participate or not and we analyze the impact of this participation on the grid by deriving equilibrium conditions~\cite{WSM000}. This analysis is done for both the rational and irrational case. For PT, we consider mainly a weighting effect.

To gain more insights on the impact of weighting on DSM participation, we consider a numerical validation using a realistic load profile in~\cite{web} which represents consumers' initial demands during Spring 2013, from the Miami International Airport. In these numerical examples, each consumer can choose a starting time to participate in DSM from the time period between $18$:$00$ and $20$:$00$. Alternatively, the consumer can decide not to participate.

\begin{figure}[!t]
 \begin{center}
 \vspace{-0.2cm}
  \includegraphics[width=8cm]{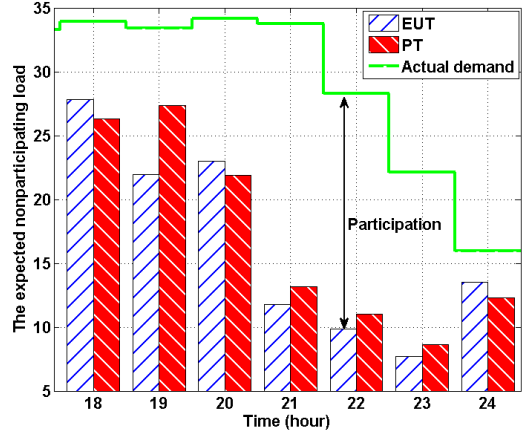}
 \vspace{-0.5cm}
   \caption{\label{fig:nonloaddifal} The expected nonparticipating load for the $6$~consumer game under both EUT and PT over $24$ hours, when consumers have different values of $\alpha$.}
\end{center}\vspace{-0.4cm}
\end{figure}

In Fig.~\ref{fig:nonloaddifal}, we show the expected nonparticipating load profile using different values for the Prelec rationality parameters $\alpha$. In this example, each consumer has a different subjective perception on other consumers and, thus, has a different rationality parameter. In particular, we choose $\alpha=[0.5\ 0.5\ 0.2\ 0.1\ 0.1\ 0.1]$ for the $6$ considered consumers.  This implies that consumer $1$ is more rational than consumers $3$-$6$ while consumers $4$-$6$ are the least rational. In this figure, we can see that, when some consumers have a very irrational observation on their opponents, the PT nonparticipating load between $21$:$00$ and $23$:$00$ will be higher for PT than EUT. This implies that, in realty, if some consumers deviate significantly from their rational strategies (for example, a consumer decides not to assist the power company in load shifting despite the economic benefit), the power company will not be able to shift the total load predicted by the rational, objective model. Thus, this simple, yet insightful example shows that one must better understand how consumers behave (here reflected by the rationality parameter) to better design the dynamic pricing and DSM scheme.

\begin{figure}[!t]
 \begin{center}
 \vspace{-0.2cm}
  \includegraphics[width=8cm]{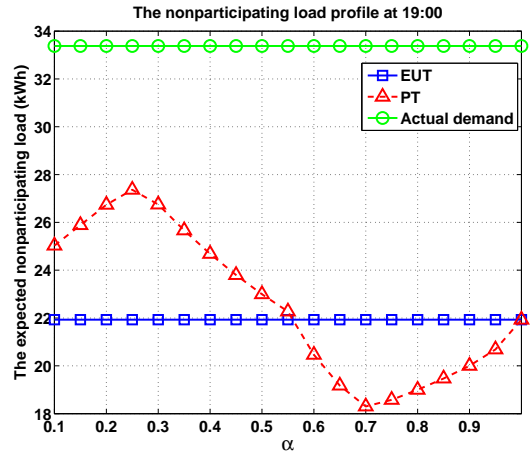}
 \vspace{-0.5cm}
   \caption{\label{fig:alpha1900} The impact of the rationality parameter $\alpha$ on the expected nonparticipating load of all consumers at $19$:$00$.}
\end{center}\vspace{-0.4cm}
\end{figure}

We further analyze the impact of the consumer rationality on DSM by showing, in Fig.~\ref{fig:alpha1900}, the expected nonparticipating load at a chosen time of the day which is here selected to be $19$:$00$ for illustrative purposes. Here, it is assumed that all consumers have a similar level of rationality. In Fig.~\ref{fig:alpha1900}, we observe that, under EUT, the expected nonparticipating load is $65.7\%$ of the total load. In contrast, under PT, the nonparticipating load is less than EUT when $\alpha>0.56$, i.e., when consumers are fairly rational.  Thus, the power company can shift more load in practice, compared to an EUT scheme, if the consumers are all of equal rationality level when $\alpha>0.56$. Clearly, there exists a rationality threshold, such that, if $\alpha$ is greater (smaller) than the threshold, PT consumers will have lower (higher) nonparticipating loads than EUT cases. A large value of $\alpha$, which maps to a small deviation from EUT, yields an increased competition thus raising the costs to the consumers. Consequently, the consumers will become risk seeking and more apt to shift their loads and decrease their payments. Thus, the increasing PT costs will force the majority to shift more loads, compared to EUT. In contrast, a relatively small rationality parameter or a large deviation from EUT, will lead to highly irrational behavior from the consumers which will lead to increasingly high competition and decreasing participation, as consumers become extremely risk averse and unwilling to participate in the DSM process.

From Fig.~\ref{fig:alpha1900}, we can infer that one of the reasons for which DSM schemes might have not been adopted widely in practice is due to a severely irrational behavior observed from the consumers. Indeed, as per Fig.~\ref{fig:alpha1900}, one can see that a small deviation from EUT (slight irrationality) may in fact be beneficial for the power company as it increases consumers' participation. In contrast, a significant deviation from EUT will inevitably lead to highly risk averse behavior which will prevent most consumers from participating; thus yielding detrimental results for the grid and preventing the power company from reaping the benefits of DSM.

Through this simple, yet realistic DSM example, we are able to see that by only considering the weighting effect of PT, the results of DSM can significantly change. This is due to the fact that the PT model allows to better capture the way in which consumers behave in practice. Consequently, this motivates a deeper investigation of the role of human decision making in practical DSM mechanisms.

\subsection{Choice of PT Parameters}
{In this section, we have brought forward several key results that show how realistic consumer behavior can impact smart grid energy management. However, in these models for assessing the rationality (e.g., the parameter $\alpha$) or risk aversion of the consumers, we have adopted PT models of risk that were conceived in the economics literature such as in ~\cite{PT00,PT01,PT02,PT03,PT04,PT05,PRELEC}. Naturally, to understand whether those models map directly into the smart grid, there is a need to run analogous behavioral experiments, with real-world smart grid consumers, to generate new empirical models for PT that can be used to further enhance the results of this section. Such experiments can be based on both qualitative surveys and on real-world simulations in which grid consumers (e.g., homeowners or factories) are solicited to participate in simulated experiments on grid scenarios that pertain to DSM, storage, or other consumer-centric features. Such experiments can mimic the gain and loss scenarios presented in Section~\ref{sec:pt2}. Using such behavioral experiments, we can refine the choice of the various PT parameters and we can generate more advanced models and results.}

\section{Conclusions and Future Outlook}\label{sec:conc}
Realizing the vision of a smart, consumer-centric grid is without any doubt strongly dependent on gathering a better understanding on the impact and role of consumer behavior in energy management processes such as demand-side management, demand response, or energy trading. In this article, we have shown that the use of prospect theory, a powerful framework from operations research and psychology, can provide the first step towards better understanding the impact of consumer behavior on smart grid operation. Indeed, our preliminary investigations have shown that consumer-related deviations from conventional, rational game-theoretic energy management mechanisms can be one of the primary reasons behind the modest adoption of such mechanisms in practical smart grid systems.

Nonetheless, in this article, we have only scratched the surface of this emerging area in smart grid research. Indeed, the study of consumer behavior in the smart grid requires significant advances to frameworks such as PT. Many future directions can be envisioned. For example, our results so far have solely relied on the analysis of the weighting effect. However, we anticipate that the use of both framing and weighting can provide deeper insights into how DSM and energy management can operate in the smart grid. Indeed, the fact that smart grid consumers will have time-dependent reference points while measuring their utility functions provides a very interesting and promising research direction.

In addition, our study thus far has focused primarily on economic-oriented models, in which the impact on the power system is restricted to load management. Instead, one can envision the use of PT-based behavioral model to better understand how the overall regulation of the power system operation can be modified due to the uncertainty and risk introduced by consumer-based decision making. Such studies can also be extended to explicitly account for communication and security considerations, both of which can involve end-user decisions.

Another important direction for future work is to explicitly account for renewable energy sources. In fact, renewable sources will introduce two types of uncertainty: a) uncertainty due to consumer decisions, as captured in the models of this paper and b) uncertainty due to nature and other environmental factors that affect renewable generation. Here, it is of interest to apply PT models to capture both types of uncertainty. Some early works on PT such as in \cite{PT05} have shown that when both weighting effects and utility uncertainty are considered, one can expect significant deviation from conventional rational results. How such deviations can be applied in a smart grid context remains an open problem.

{Moreover, the recent surge in the application of big data analytics in various smart grid scenarios will provide an important avenue to explore the differences between EUT and PT. These data that are being constantly collected can, in the future, provide an important source for corroborating the intuition provided by PT while also providing important information to derive more realistic PT models.}

{Last but not least, as mentioned previously, one important challenge is the lack of any large-scale data on how buyers and sellers will in reality behave in the still speculative smart grid market. Though PT provides broad hints about the factors that may affect choices, the tests of it have been so far restricted to basic models such as those presented here, which are still somewhat distant from the context of a large-scale practical smart grid to be determinative. In other areas, the experiments confirming PT have overwhelmingly been single session experiments in which naive participants make choices in speculative scenarios of no consequence to themselves. This is different from regular participation in a smart grid in which their choices have direct financial consequences on themselves. People have the ability to learn from experience and this is known to affect their choices in some contexts. For example, the endowment effect is that amateur collectors will not sell an item they already own for the price they would be willing to pay for it~\cite{KNETCH}. However, professional merchants do not show an endowment effect~\cite{List}. Fortunately, current internet technology makes it possible to simulate the smart grid and systematically evaluate consumer behavior in it under different conditions. Conducting such simulations will be an important area for future work as the results of such studies should make it possible to design transaction rules and human interfaces that constrain behavior into optimal pathways. Such results will also help corroborate and improve upon PT models.}

In a nutshell, the deployment of smart energy management mechanisms is an integral and essential part of the smart grid. However, in order to expedite the introduction of such features, it has become crucial to properly develop behavioral models that can factor in explicitly the impact of human behavior on the overall operation of the future, consumer-centric smart grid.

\vspace{-0.1cm}

\def\baselinestretch{1}
\bibliographystyle{IEEEtran}
\bibliography{references}

% Generated by IEEEtran.bst, version: 1.13 (2008/09/30)
\begin{thebibliography}{100}
\providecommand{\url}[1]{#1}
\csname url@samestyle\endcsname
\providecommand{\newblock}{\relax}
\providecommand{\bibinfo}[2]{#2}
\providecommand{\BIBentrySTDinterwordspacing}{\spaceskip=0pt\relax}
\providecommand{\BIBentryALTinterwordstretchfactor}{4}
\providecommand{\BIBentryALTinterwordspacing}{\spaceskip=\fontdimen2\font plus
\BIBentryALTinterwordstretchfactor\fontdimen3\font minus
  \fontdimen4\font\relax}
\providecommand{\BIBforeignlanguage}[2]{{%
\expandafter\ifx\csname l@#1\endcsname\relax
\typeout{** WARNING: IEEEtran.bst: No hyphenation pattern has been}%
\typeout{** loaded for the language `#1'. Using the pattern for}%
\typeout{** the default language instead.}%
\else
\language=\csname l@#1\endcsname
\fi
#2}}
\providecommand{\BIBdecl}{\relax}
\BIBdecl

\bibitem{SG00}
E.~Hossain, Z.~Han, and {H. V. Poor}, \emph{Smart Grid Communications and
  Networking}.\hskip 1em plus 0.5em minus 0.4em\relax Cambridge University
  Press, UK, Oct. 2012.

\bibitem{SG01}
R.~Schneiderman, ``Smart grid represents a potentially huge market for the
  electricity industry,'' \emph{IEEE Signal Processing Magazine}, vol.~27,
  no.~5, pp. 8--15, Sep. 2010.

\bibitem{SG02}
{ISO New England Inc.}, ``Overview of the smart grid: Policies, initiatives and
  needs,'' Feb. 2009.

\bibitem{FG00}
A.~Ipakchi and F.~Albuyeh, ``Grid of the future,'' \emph{IEEE Power and Energy
  Magazine}, vol.~7, no.~2, pp. 52 -- 62, Mar. 2009.

\bibitem{HAM00}
H.~Mohsenian-Rad, V.~W.~S. Wong, J.~Jatskevich, R.~Schober, and A.~Leon-Garcia,
  ``Autonomous demand side management based on game-theoretic energy
  consumption scheduling for the future smart grid,'' \emph{IEEE Trans. on
  Smart Grid}, vol.~1, no.~3, pp. 320--331, Dec. 2010.

\bibitem{DSMDR00}
Z.~Fadlallah, D.~M. Quan, N.~Kato, and I.~Stojmenovic, ``{GTES}: An optimized
  game-theoretic demand-side management scheme for smart grid,'' \emph{IEEE
  Systems Journal}, vol.~8, pp. 588--597, Jan. 2009.

\bibitem{DR02}
L.~Chen, N.~Li, S.~H. Low, and J.~C. Doyle, ``Two market models for demand
  response in power networks,'' in \emph{Proc. International Conference on
  Smart Grid Communications}, Gaithersburg, MD, USA, Oct. 2010.

\bibitem{MGA04}
S.~D. Ramchurn, P.~Vytelingum, A.~Rogers, and N.~R. Jennings, ``Agent-based
  control for decentralised demand side management in the smart grid,'' in
  \emph{Proc. International Conference on Autonomous Agents and Multiagent
  Systems (AAMAS)}, Taipei, Taiwan, May 2011.

\bibitem{QZ00}
Q.~Zhu, P.~Sauer, and T.~Ba\c{s}ar, ``Value of demand response in the smart
  grid,'' in \emph{Proc. IEEE Power and Energy Conference at Illinois},
  Champaign, IL, USA, Feb. 2013.

\bibitem{DR03}
C.~Ibars, M.~Navarro, and L.~Giupponi, ``Distributed demand management in smart
  grid with a congestion game,'' in \emph{Proc. International Conference on
  Smart Grid Communications}, Gaithersburg, MD, USA, Oct. 2010.

\bibitem{DR04}
S.~Caron and G.~Kesidis, ``Incentive-based energy consumption scheduling
  algorithms for the smart grid,'' in \emph{Proc. International Conference on
  Smart Grid Communications}, Gaithersburg, MD, USA, Oct. 2010.

\bibitem{DSMDR01}
Z.~Zhu, J.~Tang, S.~Lambotharan, W.~H. Chin, and Z.~Fan, ``An integer linear
  programming and game theory based optimization for demand-side management in
  smart grid,'' in \emph{Proc.\ IEEE Global Commun. Conf.}, Houston, TX, USA,
  Dec. 2011.

\bibitem{DSMDR02}
S.~Bu, F.~R. Yu, and P.~X. Liu, ``A game-theoretical decision-making scheme for
  electricity retailers in the smart grid with demand-side management,'' in
  \emph{Proc. International Conference on Smart Grid Communications}, Brussels,
  Belgium, Oct. 2011.

\bibitem{FF05}
S.~Bu and F.~R. Yu, ``A game-theoretical scheme in the smart grid with
  demand-side management: Towards a smart cyber-physical power
  infrastructure,'' \emph{IEEE Trans. on Emerging Topics in Computing}, vol.~1,
  pp. 22--32, Jun. 2013.

\bibitem{GT04}
H.~K. Nguyen, J.~B. Song, and Z.~Han, ``Demand side management to reduce
  peak-to-average ratio using game theory in smart grid,'' in \emph{Proc. of
  IEEE INFOCOM, Workshop on Smart Grid}, Orlando, FL, USA, Apr. 2012.

\bibitem{FF03}
S.~Mahrajan, Q.~Zhu, Y.~Zhang, S.~Gjessing, and T.~Ba\c{s}ar, ``Dependable
  demand response management in the smart grid: A stackelberg game approach,''
  \emph{IEEE Trans. on Smart Grid}, vol.~4, pp. 120--132, Mar. 2013.

\bibitem{FF04}
P.~Samadi, H.~Mohsenian-Rad, R.~Schober, and V.~W.~S. Wong, ``Advanced demand
  side management for the future smart grid using mechanism design,''
  \emph{IEEE Trans. on Smart Grid}, vol.~3, pp. 1170--1180, Sep. 2012.

\bibitem{CP04}
H.~Mohsenian-Rad and A.~Davoudi, ``Optimal demand response in {DC} distribution
  networks,'' in \emph{Proc. IEEE Int. Conf. on Smart Grid Communications
  (SmartGridComm)}, Vancouver, Canada, Oct. 2013.

\bibitem{CP00}
H.~Zhong, L.~Xie, and Q.~Xia, ``Coupon incentive-based demand response: Theory
  and case study,'' \emph{IEEE Trans. on Power Systems}, to appear 2013.

\bibitem{CP01}
M.~D. Ilic, L.~Xie, and J.~Joo, ``Efficient coordination of wind power and
  price-responsive demand {Part I}: theoretical foundations,'' \emph{IEEE
  Trans. on Power Systems}, vol.~26, pp. 1875--1884, Nov. 2011.

\bibitem{CP02}
------, ``Efficient coordination of wind power and price-responsive demand
  {Part II}: case studies,'' \emph{IEEE Trans. on Power Systems}, vol.~26, pp.
  1885--1893, Nov. 2011.

\bibitem{CP08}
C.~Su and D.~Kirschen, ``Quantifying the effect of demand response on
  electricity markets,'' \emph{IEEE Trans. on Power Systems}, vol.~24, pp.
  1199--1207, Aug. 2009.

\bibitem{DSMDR03}
P.~Palensky and D.~Dietrich, ``Demand side management: Demand response,
  intelligent energy systems, and smart loads,'' \emph{IEEE Trans. on
  Industrial Informatics}, vol.~7, no.~3, pp. 381 -- 388, Aug. 2011.

\bibitem{DSMDR04}
B.~Asare-Bediako, W.~L. Kling, and P.~F. Ribeiro, ``Integrated agent-based home
  energy management system for smart grids applications,'' in \emph{Proc.
  IEEE/PES Innovative Smart Grid Technologies Europe}, Lyngby, Denmark, Oct.
  2013.

\bibitem{DSMDR05}
H.~Mohsenian-rad, V.~W.~S. Wong, J.~Jatskevich, and R.~Schober, ``Optimal and
  autonomous incentive-based energy consumption scheduling algorithm for smart
  grid,'' in \emph{Proc. Innovative Smart Grid Technologies (ISGT)},
  Gathersburg, MD, USA, Jan. 2010.

\bibitem{DSMDR06}
J.~Ma, J.~Deng, L.~Song, and Z.~Han, ``Incentive mechanism for demand side
  management in smart grid using auction,'' \emph{IEEE Trans. on Smart Grid},
  vol.~5, no.~3, pp. 1379 -- 1388, May 2014.

\bibitem{DSMDR07}
Y.~Liu, C.~Yuen, S.~Huang, N.~Hassan, X.~Wang, and S.~Xie, ``Peak-to-average
  ratio constrained demand-side management with consumer's preference in
  residential smart grid,'' \emph{IEEE Journal on Selected Topics in Signal
  Processing}, vol.~8, no.~6, pp. 1084 -- 1097, Dec. 2014.

\bibitem{DSMDR08}
Z.~M. Fadlullah, M.~Q. Dong, N.~Kato, and I.~Stojmenovic, ``A novel game-based
  demand side management scheme for smart grid,'' in \emph{Proc.\ IEEE Wireless
  Commun. and Networking Conf.}, Shanghai, China, Apr. 2013.

\bibitem{DSMDR09}
C.~O. Adika and L.~Wang, ``Smart charging and appliance scheduling approaches
  to demand side management,'' \emph{International Journal of Electrical Power
  \& Energy Systems}, vol.~57, pp. 232--240, May 2014.

\bibitem{DSMDR10}
J.~Yang, G.~Zhang, and K.~Ma, ``Real-time pricing-based scheduling strategy in
  smart grids: A hierarchical game approach,'' \emph{Journal of Applied
  Mathematics}, vol. 2014, Apr. 2014.

\bibitem{DSMDR12}
X.~Xue, S.~Wang, C.~Yan, and B.~Cui, ``A fast chiller power demand response
  control strategy for buildings connected to smart grid,'' \emph{Applied
  Energy}, vol. 137, p. 77–87, Jan. 2015.

\bibitem{DSMDR13}
F.~Rahimi and A.~Ipakchi, ``Demand response as a market resource under the
  smart grid paradigm,'' \emph{IEEE Trans. Smart Grid}, vol.~1, no.~1, pp.
  82--88, Apr. 2010.

\bibitem{DSMDR14}
Q.~Zhu, Z.~Han, and T.~Ba\c{s}ar, ``A differential game approach to distributed
  demand side management in smart grid,'' in \emph{Proc.\ Int.\ Conf.\ on
  Communications}, Ottawa, Canada, Jun. 2012.

\bibitem{DSMDR15}
E.~Nekouei, T.~Alpcan, and D.~Chattopadhyay, ``A game-theoretic analysis of
  demand response in electricity markets,'' in \emph{Proc. IEEE PES General
  Meeting}, National Harbor, MD, USA, Jul. 2014.

\bibitem{DSMDR16}
C.~Joe-Wong, S.~Sen, S.~Ha, and M.~Chiang, ``Optimized day-ahead pricing for
  smart grids with device-specific scheduling flexibility,'' \emph{{IEEE} J.
  Select. Areas Commun.}, vol.~30, no.~6, pp. 1075 -- 1085, Jul. 2012.

\bibitem{DSMDR17}
A.~Safdarian, M.~Fotuhi-Firuzabad, and M.~Lehtonen, ``A distributed algorithm
  for managing residential demand response in smart grids,'' \emph{IEEE Trans.
  on Industrial Informatics}, vol.~10, no.~4, pp. 2385 -- 2393, Nov. 2014.

\bibitem{DSMDR18}
T.~Logenthiran, D.~Srinivasan, and K.~W.~M. Vanessa, ``Demand side management
  of smart grid: Load shifting and incentives,'' \emph{AIP Journal of Renewable
  and Sustainable Energy}, vol.~6, Jun. 2014.

\bibitem{MGA}
P.~Vytelingum, T.~D. Voice, S.~D. Ramchurn, A.~Rogers, and N.~R. Jennings,
  ``Agent-based micro-storage management for the smart grid,'' in \emph{Proc.
  International Conference on Autonomous Agents and Multiagent Systems
  (AAMAS)}, Toronto, Canada, May 2010.

\bibitem{GT06}
I.~Atzeni, L.~G. Ordonez, G.~Scutari, D.~P. Palomar, and J.~R. Fonollosa,
  ``Noncooperative and cooperative optimization of distributed energy
  generation and storage in the demand-side of the smart grid,'' \emph{{IEEE}
  Trans. Signal Processing}, vol.~61, no.~10, pp. 2454--2472, May 2013.

\bibitem{HAM01}
C.~Wu, H.~Mohsenian-Rad, and J.~Huang, ``Vehicle-to-aggregator interaction
  game,'' \emph{IEEE Trans. on Smart Grid}, vol.~3, no.~1, pp. 434 -- 442, Oct.
  2011.

\bibitem{WI03}
Y.~Wang, W.~Saad, Z.~Han, H.~V. Poor, and T.~Ba\c{s}ar, ``A game-theoretic
  approach to energy trading in the smart grid,'' \emph{IEEE Trans. on Smart
  Grid}, vol.~5, no.~3, pp. 1439--1450, Apr. 2014.

\bibitem{REN00}
C.~Cecati, C.~Citro, and P.~Siano, ``Combined operations of renewable energy
  systems and responsive demand in a smart grid,'' \emph{IEEE Trans. on
  Sustainable Energy}, vol.~3, no.~1, pp. 468 -- 476, Jul. 2011.

\bibitem{STOR00}
H.~M. Soliman and A.~Leon-Garcia, ``Game-theoretic demand-side management with
  storage devices for the future smart grid,'' \emph{IEEE Trans. on Smart
  Grid}, vol.~5, no.~3, pp. 1475 -- 1485, May 2014.

\bibitem{STOR01}
Z.~Fan, ``A distributed demand response algorithm and its application to phev
  charging in smart grids,'' \emph{IEEE Trans. Smart Grid}, vol.~3, no.~3, pp.
  1280 -- 1290, Sep. 2012.

\bibitem{STOR02}
M.~A. Lopez, S.~{de la Torre}, S.~Martin, and J.~A. Aguado, ``Demand-side
  management in smart grid operation considering electric vehicles load
  shifting and vehicle-to-grid support,'' \emph{International Journal of
  Electrical Power and Energy Systems}, vol.~64, p. 689–698, Jan. 2015.

\bibitem{STOR03}
W.~Tushar, W.~Saad, H.~V. Poor, and D.~B. Smith, ``Economics of electric
  vehicle charging: A game theoretic approach,'' \emph{IEEE Trans. on Smart
  Grid}, vol.~3, no.~4, pp. 1767--1779, Dec. 2012.

\bibitem{STOR04}
I.~Atzeni, L.~G. Ordonez, G.~Scutari, D.~P. Palomar, and J.~R. Fonollosa,
  ``Demand-side management via distributed energy generation and storage
  optimization,'' \emph{IEEE Trans. on Smart Grid}, vol.~4, no.~2, pp. 866 --
  876, Jun. 2013.

\bibitem{STOR05}
S.~Lakshminarayana, T.~Q.~S. Quek, and H.~V. Poor, ``Combining cooperation and
  storage for the integration of renewable energy in smart grids,'' in
  \emph{Proc. of IEEE INFOCOM, Workshop on Smart Grid}, Toronto, Canada, Apr.
  2014.

\bibitem{STOR06}
Q.~Sun, M.~E. Cotterell, A.~Beach, and S.~Grijalva, ``The fundamental value of
  information and strategy in stochastic management of distributed energy
  storage,'' in \emph{Proc. North American Power Symposium}, Champaign, IL,
  USA, Sep. 2012.

\bibitem{STOR07}
M.~Ampatzis, P.~H. Nguyen, and W.~L. Kling, ``Introduction of storage
  integrated pv sytems as an enabling technology for smart energy grids,'' in
  \emph{Proc. Innovative Smart Grid Technologies (ISGT) Europe}, Lyngby,
  Denmark, Oct. 2013.

\bibitem{EW01}
R.~Couillet, S.~M. Perlaza, H.~Tembine, and M.~Debbah, ``Electrical vehicles in
  the smart grid: A mean field game analysis,'' \emph{{IEEE} J. Select. Areas
  Commun.}, vol.~30, no.~6, pp. 1086 -- 1096, Jul. 2012.

\bibitem{EW07}
N.~Rotering and M.~Ilic, ``Optimal plug-in electric vehicle charge control in
  deregulated electricity markets,'' \emph{IEEE Transactions on Power Systems},
  vol.~26, no.~3, pp. 1021 -- 1029, Aug. 2011.

\bibitem{EW08}
C.~Silva, M.~Ross, and T.~Farias, ``Evaluation of energy consumption, emissions
  and cost of plug-in hybrid vehicles,'' \emph{Elsevier Energy Conversion and
  Management}, vol.~50, no.~7, pp. 1635--1643, Jul. 2009.

\bibitem{EW02}
S.~Sojoudi and S.~H. Low, ``Optimal charging of plug-in hybrid electric
  vehicles in smart grids,'' \emph{IEEE Power and Energy Society (PES) General
  Meeting}, Jul. 2011.

\bibitem{EW05}
E.~Tara, S.~Shahidinejad, S.~Filizadeh, and E.~Bibeau, ``Battery storage sizing
  in a retrofitted plug-in hybrid electric vehicle,'' \emph{IEEE Transactions
  on Vehicular Technology}, vol.~59, no.~6, pp. 2786--2794, Jul. 2010.

\bibitem{EW06}
J.~Gonder, T.~Markel, M.~Thornton, and A.~Simpson, ``Using global positioning
  system travel data to assess real-world energy use of plug-in hybrid electric
  vehicles,'' \emph{Transportation Research Record: Journal of the
  Transportation Research Board}, vol. 2007, no. 2017, pp. 26--32, Jan. 2008.

\bibitem{ET00}
T.~A. Al-Awami and E.~Sortomme, ``Coordinating vehicle-to-grid services with
  energy trading,'' \emph{IEEE Trans. Smart Grid}, vol.~3, no.~1, pp. 453 --
  462, Mar. 2012.

\bibitem{ET01}
D.~Ilic, P.~G. {Da Silva}, S.~Karnouskos, and M.~Griesemer, ``An energy market
  for trading electricity in smart grid neighbourhoods,'' in \emph{Proc. 6th
  IEEE Int. Conference on Digital Ecosystems Technologies}, Campione D'Italia,
  Italy, Jun. 2012.

\bibitem{ET02}
B.-G. Kim, S.~Ren, M.~{van der Schaar}, and J.-W. Lee, ``Bidirectional energy
  trading and residential load scheduling with electric vehicles in the smart
  grid,'' \emph{{IEEE} J. Select. Areas Commun.}, vol.~31, no.~7, pp. 1219 --
  1234, Jul. 2013.

\bibitem{ET03}
W.~Tushar, J.~A. Zhang, D.~B. Smith, H.~V. Poor, and S.~Thiebaux,
  ``Prioritizing consumers in smart grid: A game theoretic approach,''
  \emph{IEEE Trans. on Smart Grid}, vol.~5, no.~3, pp. 1429 -- 1438, May 2014.

\bibitem{ET04}
S.~Chen, N.~B. Shroff, and P.~Sinha, ``Energy trading in the smart grid: From
  end-user's perspective,'' in \emph{Proc. Asilomar Conference on Signals,
  Systems, and Computers}, Pacific Grove, CA, USA, Nov. 2013.

\bibitem{ET05}
S.~Chen, N.~Shroff, and P.~Sinha, ``Heterogeneous delay tolerant task
  scheduling and energy management in the smart grid with renewable energy,''
  \emph{{IEEE} J. Select. Areas Commun.}, vol.~31, no.~7, pp. 1258 -- 1267,
  Jul. 2013.

\bibitem{ET06}
I.~S. Bayram, M.~Z. Shakir, M.~Abdallah, and K.~Qaraqe, ``A survey on energy
  trading in smart grid,'' in \emph{Proc. IEEE IEEE Global Conference on Signal
  and Information Processing (GlobalSIP)}, Atlanta, GA, USA, Dec. 2014.

\bibitem{ET07}
D.~T. Nguyen and L.~Le, ``Optimal energy trading for building microgrid with
  electric vehicles and renewable energy resources,'' in \emph{Proc. IEEE PES
  Innovative Smart Grid Technologies (ISGT)}, Washington, DC, USA, Feb. 2014.

\bibitem{ET08}
E.~Mocanu, K.~O. Aduda, P.~H. Nguyen, G.~Boxem, W.~Zeiler, M.~Gibescu, and
  W.~L. Kling, ``Optimizing the energy exchange between the smart grid and
  building systems,'' in \emph{Proc. 49th Int. Universities Power Engineering
  Conference}, Cluj-Napoca, Romania, Sep. 2014.

\bibitem{ET09}
A.~Mondal and S.~Misra, ``Dynamic coalition formation in a smart grid: A game
  theoretic approach,'' in \emph{Proc.\ Int.\ Conf.\ on Communications},
  Budapest, Hungary, Jun. 2013.

\bibitem{FAHEY}
J.~Fahey, ``Companies dangle free nights and weekends,'' \emph{Pocono Record},
  2013.

\bibitem{FAHEY2}
P.~Durand, ``Moving beyond smart grid toward customer engagement,''
  \emph{Electric Light \& Power}, Jun. 2015.

\bibitem{FAHEY3}
G.~Wang, ``3 ways to engage utility customers in home energy management,''
  \emph{Electric Light \& Power}, Jun. 2015.

\bibitem{FAHEY4}
{Associated Press}, ``Small-scale solar power market draws big utilities,''
  Oct. 2015.

\bibitem{FAHEY6}
\BIBentryALTinterwordspacing
{Solar Energy Industries Association}, ``Solar energy facts: 2014 year in
  review,'' 2015. [Online]. Available:
  \url{http://www.seia.org/research-resources/solar-industry-data}
\BIBentrySTDinterwordspacing

\bibitem{FAHEY5}
D.~Cardwell, ``Compromise in arizona defers a solar power fight,'' \emph{New
  York Times}, 2015.

\bibitem{FAHEY7}
\BIBentryALTinterwordspacing
{Federal Energy Regulatory Commission}, ``Demand response \& advance metering
  staff report,'' 2012. [Online]. Available:
  \url{https://www.ferc.gov/legal/staff-reports/12-20-12-demand-response.pdf}
\BIBentrySTDinterwordspacing

\bibitem{GTBOOK}
T.~Ba\c{s}ar and G.~J. Olsder, \emph{Dynamic Noncooperative Game Theory}.\hskip
  1em plus 0.5em minus 0.4em\relax Philadelphia, PA, USA: SIAM Series in
  Classics in Applied Mathematics, 1999.

\bibitem{MIC00}
M.~E. Khodayar, M.~Barati, and M.~Shahidehpour, ``Integration of high
  reliability distribution system in microgrid operation,'' \emph{IEEE Trans.
  on Smart Grid}, vol.~3, no.~4, pp. 1997 -- 2006, Dec. 2012.

\bibitem{MIC01}
P.~Piagi and R.~H. Lasseter, ``Autonomous control of microgrids,'' \emph{IEEE
  Power Engineering Society General Meeting}, vol.~27, no.~5, pp. 78--94, Oct.
  2006.

\bibitem{MIC02}
F.~Katiraei, R.~Iravani, and N.~Hatziargyriou, ``Microgrids management,''
  \emph{IEEE Power and Energy Magazine}, vol.~6, pp. 54--65, May 2008.

\bibitem{MIC03}
N.~Hatziargyriou, H.~Sano, R.~Iravani, and C.~Marnay, ``Microgrids: An overview
  of ongoing research development and demonstration projects,'' \emph{IEEE
  Power and Energy Magazine}, vol.~27, pp. 78--94, Aug. 2007.

\bibitem{MIC04}
S.~Dasgupta, S.~N. Mohan, S.~K. Sahoo, and S.~K. Panda, ``Lyapunov
  function-based current controller to control active and reactive power flow
  from a renewable energy source to a generalized three-phase microgrid
  system,'' \emph{IEEE Trans. on Industrial Electronics}, vol.~60, no.~2, pp.
  799 -- 813, Feb. 2013.

\bibitem{MIC05}
A.~Karabibera, C.~Kelesb, A.~Kaygusuzb, and B.~B. Alagoz, ``An approach for the
  integration of renewable distributed generation in hybrid {DC/AC}
  microgrids,'' \emph{Renewable Energy}, vol.~51, p. 251–259, Apr. 2013.

\bibitem{MIC06}
O.~Hafez and K.~Bhattacharya, ``Optimal planning and design of a renewable
  energy based supply system for microgrids,'' \emph{Renewable Energy},
  vol.~45, p. 7–15, Sep. 2012.

\bibitem{MIC07}
M.~Sechilariu, B.~Wang, and F.~Locment, ``Building integrated photovoltaic
  system with energy storage and smart grid communication,'' \emph{IEEE Trans.
  on Industrial Electronics}, vol.~60, no.~4, pp. 1607 -- 1618, Apr. 2013.

\bibitem{MIC08}
C.~A. Hill, M.~Such, C.~D, J.~Gonzalez, and W.~M. Grady, ``Battery energy
  storage for enabling integration of distributed solar power generation,''
  \emph{IEEE Trans. on Smart Grid}, vol.~3, no.~2, pp. 850 -- 857, May 2012.

\bibitem{HP05}
L.~Xie, D.-H. Choi, S.~Kar, and H.~V. Poor, ``Fully distributed state
  estimation for wide-area monitoring systems,'' \emph{IEEE Trans. on Smart
  Grid}, vol.~3, no.~3, pp. 1154--1169, Sep. 2012.

\bibitem{MOMA00}
\BIBentryALTinterwordspacing
{Modellstadt Mannheim}, ``E-energy-projectsmart city mannheim,''
  \emph{Technical Report}, 2012. [Online]. Available:
  \url{http://ec.europa.eu/dgs/jrc/downloads/events/20120711-esof/esof-2012-thomas-wolski.pdf}
\BIBentrySTDinterwordspacing

\bibitem{MOMA01}
\BIBentryALTinterwordspacing
{Power Plus Communications}, ``Smart citizens,'' \emph{Modellstadt Mannheim
  Technical Report}, 2012. [Online]. Available:
  \url{http://www.ppc-ag.de/files/moma_en.pdf}
\BIBentrySTDinterwordspacing

\bibitem{NREL}
{EnerNex Corp.}, ``Eastern wind integration and transmission study,''
  \emph{National Renewable Energy Laborator, Report NREL/SR-550-47078}, 2010.

\bibitem{NREL1}
\BIBentryALTinterwordspacing
{OpenEI}, ``Open energy data sets,'' 2014. [Online]. Available:
  \url{http://en.openei.org/datasets/}
\BIBentrySTDinterwordspacing

\bibitem{NREL2}
\BIBentryALTinterwordspacing
{U.S. Energy Information Administration}, ``Electricity statistics and data,''
  2014. [Online]. Available: \url{http://www.eia.gov/electricity/data.cfm}
\BIBentrySTDinterwordspacing

\bibitem{NREL3}
\BIBentryALTinterwordspacing
{ecoEnergy}, ``Energy consumption of major household appliances shipped in
  canada,'' \emph{Technical Report}, Dec. 2007. [Online]. Available:
  \url{http://oee.nrcan.gc.ca/Publications/statistics/cama07/pdf/cama07.pdf}
\BIBentrySTDinterwordspacing

\bibitem{NREL4}
\BIBentryALTinterwordspacing
{Lawrence Berkeley National Laboratory}, ``Standy power data,'' 2014. [Online].
  Available: \url{http://standby.lbl.gov/data.html}
\BIBentrySTDinterwordspacing

\bibitem{NREL5}
\BIBentryALTinterwordspacing
{Massachusetts Institute of Technology}, ``The reference energy disaggregation
  data set (redd),'' 2014. [Online]. Available:
  \url{http://redd.csail.mit.edu/}
\BIBentrySTDinterwordspacing

\bibitem{PT00}
D.~Kahneman and A.~Tversky, ``Prospect theory: An analysis of decision under
  risk,'' \emph{Econometrica}, vol.~47, pp. 263--291, 1979.

\bibitem{PT01}
G.~A. Quattrone and A.~Tversky, ``Contrasting rational and psychological
  analyses of political choice,'' \emph{The American Political Science Review},
  vol.~82, no.~3, pp. 719--736, 1988.

\bibitem{PT02}
C.~Camerer, L.~Babcock, G.~Loewenstein, and R.~Thaler, ``Labor supply of {New
  York City} cab drivers: One day at a time,'' \emph{Quarterly Journal of
  Economics}, no. 111, pp. 408--441, May 1997.

\bibitem{PT03}
D.~Kahneman and A.~Tversky, \emph{Choices, Values, and Frames}.\hskip 1em plus
  0.5em minus 0.4em\relax Cambridge University Press, 2000.

\bibitem{PT04}
Y.~Wang, A.~Nakao, and J.~Ma, ``Psychological research and application in
  autonomous networks and systems: A new interesting field,'' in \emph{Proc.
  International Conference on Intelligent Computing and Integrated Systems},
  Guilin, China, Oct. 2010.

\bibitem{PT05}
\BIBentryALTinterwordspacing
L.~P. Metzger and M.~O. Riegery, ``Equilibria in games with prospect theory
  preferences,'' \emph{Working Paper}, Nov. 2009. [Online]. Available:
  \url{http://www.bf.uzh.ch/publikationen/pdf/publ_2150.pdf}
\BIBentrySTDinterwordspacing

\bibitem{NM00}
T.~Li and N.~Mandayam, ``Prospects in a wireless random access game,'' in
  \emph{Proc. 46th Annual Conference on Information Sciences and Systems},
  Princeton, NJ, USA, Mar. 2012.

\bibitem{NM01}
------, ``When users interfere with protocols: prospect theory in wireles
  networks using random access as an example,'' \emph{{IEEE} Trans. Wireless
  Commun.}, vol.~13, no.~4, pp. 1888--1907, Feb. 2014.

\bibitem{CPT00}
A.~Tversky and D.~Kahneman, ``Advances in prospect theory: Cumulative
  representation of uncertainty,'' \emph{Journal of Risk and Uncertainty},
  vol.~5, pp. 297--323, Oct. 1992.

\bibitem{KAH00}
D.~Kahneman, \emph{Thinking, fast and slow}.\hskip 1em plus 0.5em minus
  0.4em\relax New York City, NY, USA: Farrar, Straus, \& Giroux, 2011.

\bibitem{PRELEC}
D.~Prelec, ``The probability weighting function,'' \emph{Econometrica}, pp.
  497--528, 1998.

\bibitem{WSM00}
Y.~Wang, W.~Saad, N.~Mandayam, and H.~V. Poor, ``Integrating energy storage in
  the smart grid: A prospect-theoretic approach,'' in \emph{Proc.\ IEEE Int.\
  Conf.\ on Acoustics, Speech, and Signal Processing (ICASSP)}, Florence,
  Italy, May 2014.

\bibitem{YW00}
Y.~Wang and W.~Saad, ``On the role of utility framing in smart grid energy
  storage management,'' in \emph{Proc.\ Int.\ Conf.\ on Communications,
  Workshop on Smart Grids}, London, UK, Jun. 2015.

\bibitem{WSM000}
Y.~Wang, W.~Saad, N.~Mandayam, and H.~V. Poor, ``Load shifting in the smart
  grid: To participate or not?'' \emph{IEEE Trans. on Smart Grid}, to appear
  2015.

\bibitem{web}
``Open{EI}, \url{http://en.openei.org/datasets/files/961/pub/}.''

\bibitem{KNETCH}
J.~L. Knetsch, ``The endowment effect and evidence of nonreversible
  indifference curves,'' \emph{American Economic Review}, vol.~79, pp.
  1277--1284, 1989.

\bibitem{List}
J.~A. List, ``Does market experience eliminate market anomalies?''
  \emph{Quarterly Journal of Economics}, vol. 118, pp. 47--71, 2003.

\end{thebibliography}

% that's all folks
\end{document}